\documentclass[12pt]{article}
\usepackage{amsfonts}
\usepackage{amssymb}
\usepackage{amscd}

\def\a{\alpha}
\def\b{\beta}
\def\g{\gamma}
\def\d{\delta}

\def\e{\epsilon}

\def\th{\theta}
\def\k{\kappa}
\def\l{\lambda}

\def\o{\omega}

\def\c{\chi}

\newcommand{\cl}{{\mathcal L}}

\newcommand{\C}{\mathbb C}
\newcommand{\R}{\mathbb R}

\newcommand{\Z}{\mathbb Z}

\def\ba{{\bar{a}}}

\def\bx{{\bar{x}}}

\def\tj{{\tilde{j}}}

\def\tm{{\tilde{m}}}

\def\tpsi{{\widetilde\Psi}}
\def\tphi{{\widetilde\Phi}}

\newcommand{\SDiff}{{\mathop{\mbox{SDiff}}\nolimits\,}}
\newcommand{\LSDiff}{{\mathop{\mbox{LSDiff}}\nolimits\,}}

\def\p{\partial}
\def\pa{\partial}
\def\dt#1{{\buildrel {\hbox{\LARGE .}} \over {#1}}}  
\def\ad{{\dt{\alpha}}}
\def\bd{{\dt{\beta}}}
\def\gd{{\dt{\gamma}}}
\def\dd{{\dt{\delta}}}

\def\sfrac#1#2{{\textstyle\frac#1#2}}
\def\cor{\widehat=}
\def\tr{{\rm tr}}

\def\be{\begin{equation}}
\def\ee{\end{equation}}
\def\bea{\begin{eqnarray}}
\def\eea{\end{eqnarray}}
\def\gl#1{(\ref{#1})}

\begin{document}
\begin{titlepage}
\begin{flushright}
hep-th/9912281\\
ITP--UH--37/99 \\
December, 1999
\end{flushright}

\vskip 2.0cm

\begin{center}
{\Large\bf Mathematics and Physics of N=2 Strings}

\vskip 1.5cm

{\Large \ Olaf Lechtenfeld}

\vskip 0.5cm

{\it Institut f\"ur Theoretische Physik, Universit\"at Hannover}\\
{\it Appelstra\ss{}e 2, 30167 Hannover, Germany}\\
{URL: www.itp.uni-hannover.de/\~{}lechtenf}

\end{center}
\vskip 1.5cm

\begin{abstract}

Open and closed strings with two worldsheet supersymmetries in
$2{+}2$ dimensional spacetime are reviewed in the NSR formulation.
I briefly discuss their quantization, mutual and self-interactions, 
classical spacetime dynamics and interpretation in terms of self-dual
Yang-Mills and gravity. A stringy origin of the infinite self-dual 
gravity hierarchy is presented. An outlook to the loop level concludes.

\end{abstract}

\vfill

Talk presented at SQS-99 during 27--31 July, 1999, at JINR, Dubna, Russia

\end{titlepage}

\newpage

\bigskip\noindent
{\bf 1.\ \ Introduction.}
\smallskip

Strings and point particles are intimately connected.
The most direct relation appears in the infrared limit,
where the string tension~$T$ blows up and only the massless
string modes (i.e. particles) survive with their characteristic dynamics:
Gauge theory and gravity, with or without supersymmetry,
emerge at leading order in $1/T$ as their effective spacetime description.

\medskip

The five 10d superstring theories, together with 11d supergravity,
are connected within the framework of M theory. Yet, besides the
(inconsistent) bosonic string, there exists still another kind of
string theory, namely those possessing {\it two\/} worldsheet supersymmetries
and being critical in $2{+}2$ dimensions.
The understanding of these so-called $N{=}2$ strings and their spacetime 
dynamics was critically enhanced by the work of Ooguri and Vafa~\cite{OV}.
They discovered that the single massless scalar mode present in such strings 
again encodes Yang-Mills or gravity, but with a self-duality restriction 
rather than a supersymmetric extension!

\medskip

Over the last ten years, further progress has been made in these matters,
leading to fascinating interrelations between the mathematical structures
and the physical aspects of this strange kind of strings.

\bigskip\noindent
{\bf 2.\ \ N=2 Strings -- Quantization.}
\smallskip

{}From the worldsheet point of view, critical $N{=}2$ strings
in flat Kleinian space $\R^{2,2}$ 
are a theory of $N{=}2$ supergravity $(h,\c,A)$ on a $1{+}1$ dimensional
Riemann surface (with punctures and possibly boundaries),
coupled to two chiral $N{=}2$ massless matter multiplets $(x,\psi)$.
The latter's components are complex scalars (the four string coordinates)
and $SO(1,1)$ Dirac spinors (their four NSR partners).
The Brink-Schwarz worldsheet action~\cite{BS} enjoys 
$N{=}2$ super coordinate and Weyl invariance on the worldsheet, 
as well as global $U(1,1)$ target spacetime symmetry. 
Hence, as a gauge theory, it represents a constraint system.
As is well known, those constraints are in fact needed to get rid of 
the negative norm Fock states inevitable in any relativistic theory
with spin one and higher. Physical states are required to obey
\ (constraints)$|{\rm phys}\rangle=0$, and exactly for $D{=}2{+}2$ 
this condition forecloses negative norm states and decouples all zero
norm states, yielding $\ \langle{\rm phys}|{\rm phys}\rangle\ge0$.

\medskip

The most appropriate technique to covariantly quantize gauge theories
makes use of the cohomology of the BRST operator~$Q$, which acts on a 
ghost-extended Fock space of string states. 
In this way, one finds that the spectrum of physical states is completely
devoid of the excited (massive) states which are the hallmark of strings.
Merely a single massless scalar state $|k\rangle$ at the lowest mass level,
$k^2{=}0$ (no tachyons either), appears, which in the open string case
is dressed with Chan-Paton labels: $|k,A\rangle$.
In particular, there is no room for spacetime fermions or supersymmetry.
Thus, $N{=}2$ strings seem to be identical to their effective spacetime
field theories, which contain a $U(1,1)$ scalar field, named $\Psi(x)$
in the closed-string case and $\Phi(x)$ (Lie algebra valued) in the
open-string case~\cite{OV}.

\medskip

Naively, one would expect a string theory in $\R^{2,2}$ to possess
global $SO(2,2)$ `Lorentz' symmetry. However, the necessity to introduce
a complex structure ($\R^{2,2}\to\C^{1,1}$) reduces the invariance,
\be \label{break1}
{\rm Spin}(2,2)\ =\ SU(1,1) \times SU(1,1)' \
\longrightarrow\ U(1) \times SU(1,1)' \ \simeq\ U(1,1) \quad,
\ee
making it natural to use a complex index notation,
$(x)=(x^a,\bx^\ba)$ with $a\in\{1,2\}$.
The moduli space of complex structures is the two-sheeted hyperboloid
$H^2=H_+^2\cup H_-^2$ with $H_\pm^2\simeq SU(1,1)/U(1)$.
It can be completed to $CP^1$ by sewing the two sheets together along
a circle, $CP^1=H_+^2 \cup S^1 \cup H_-^2$.

\medskip

Interestingly, the angle $\th$ of this $S^1$ can be interpreted as the
Maxwell instanton angle appearing as one of two string coupling constants.
More precisely, the perturbative expansion in worldsheet topologies
gives string configurations with $U(1)$ (Maxwell) bundles of Chern number
(instanton number) $M:=\frac{1}{2\pi}\smallint F$ (with $F{=}dA$ locally)
a weight factor of $e^{iM\th}$.
In addition, the angle $\th$ characterizes a real polarization or cotangent
structure, $T\R^{2,2}\to\R^2\oplus\R^2$, which defines a pair of null planes.
In this context, 
it proves convenient to employ a real $SL(2,\R)$ spinor index notation,
$(x)=(x^{\a\ad})$ with $\a\in\{+,-\}$,
and each null vector factorizes into two real spinors, 
$u_{\a\ad}=v_\a w_\ad$.
The spinor~$v$, which (modulo scale) determines the null planes,
now happens to encode not only the instanton angle $\th$ in its phase
but also the other (open) string coupling~$g\in\R^+$, 
weighing the Euler number of the worldsheet, in its modulus~\cite{BL1}: 
\be\label{spinor}
{v_+\choose v_-}\ =\ \sqrt{g}\,{\cos\sfrac\th2 \choose \sin\sfrac\th2} \quad.
\ee
Therefore, 
the space of string couplings is that of nonzero real $SL(2,\R)$ spinors,
and fixing their values amounts to breaking the global `Lorentz' invariance
of $\R^{2,2}$ in a way different from \gl{break1},
\be \label{break2}
{\rm Spin}(2,2)\ =\ SL(2,\R) \times SL(2,\R)' \
\longrightarrow\ \R \times SL(2,\R)' \quad,
\ee
where $\R$ here stands for the stability group of the spinor~$v$,
\be
\Biggl\{ 
{1{-}\lambda\sin\sfrac\th2\cos\sfrac\th2 \qquad \lambda\cos^2\sfrac\th2\quad 
\choose 
\quad-\lambda\sin^2\sfrac\th2 \qquad 1{+}\lambda\sin\sfrac\th2\cos\sfrac\th2\,}
\Biggm|\ \lambda\in\R \Biggr\} \quad.
\ee

\medskip

In place of the Euler number I shall use the `spin' variable
\be
J\ =\ 2n_c+n_o-4+4(\#{\rm handles})+2(\#{\rm boundaries})+2(\#{\rm crosscaps})
\ \in\ \Z 
\ee
for worldsheets of a given topology, 
with $n_c$ closed and $n_o$ open string legs.
While the open string coupling~$g$ as a gauge coupling is dimensionless 
in $2{+}2$ dimensions, the closed string coupling~$\k$, being a
gravitational coupling, carries dimension $[\k]=$ cm.
Their relation is $\k=g^2/\sqrt{T}$ where $T$ is the string tension.
Since I argue that the string couplings $(g,e^{i\th})$
can be changed at will by global $SO(2,2)$ transformations not preserving~$v$,
it is admissible to rotate them to $(g,e^{i\th}){=}(1,1)$ 
by a convenient choice of Lorentz frame, and I will do so for most
of this talk. In this fashion, only the highest weights of $SL(2,\R)$
representations occur.

\bigskip\noindent
{\bf 3.\ \ N=2 Strings -- Interactions.}
\smallskip

So far, I have extracted from $N{=}2$ string theory a scalar field 
($\Psi$ or $\Phi$) with a massless dispersion. To find its dynamics, 
one must compute its self-interaction, 
which is determined perturbatively by the string scattering amplitudes.
The first-quantized string path integral for the $(n_c,n_o)$-point
function $A^{(n_c,n_o)}$ includes a sum over worldsheet topologies~$(J,M)$,
weighted with appropriate powers in the string couplings~$(g,e^{i\th})$: 
\be\label{amp}
A^{(n_c,n_o)}(g,\th) 
= \!\!\!\sum_{J\ge2n_c+n_o-4}\!\! g^J 
\sum_{M=-J}^{+J} {\textstyle{2J\choose J{+}M}}\,
\sin^{J-M}\sfrac\th2\,\cos^{J+M}\sfrac\th2\,A^{(n_c,n_o)}_{J,M}
\ee
where the instanton sum has a finite range because bundles with $|M|{>}J$
do not contribute.
All length dimensions (i.e. powers of $T$) have been absorbed into the
partial amplitudes $A^{(n_c,n_o)}_{J,M}$, which are integrals over the metric,
gravitini, and Maxwell moduli spaces. The integrands may be obtained
as correlation functions of vertex operators in the $N{=}(2,2)$ superconformal
field theory on the worldsheet surface of fixed shape (moduli) and
topology.

\medskip

In momentum space and for the frame $(g,\th){=}(1,0)$,
\be
A^{(n_c,n_o)}(k_1,\ldots,k_n)\ =\
\d_{k_1{+}\ldots{+}k_n,0}\,\sum_J \tilde{A}^{(n_c,n_o)}_{J,J}(k_1,\ldots,k_n)
\ee
with $n:=n_c{+}n_o$,
and $\tilde{A}^{(n_c,n_o)}_{J,J}$ should be identified as the contribution 
of a class of Feynman diagrams (belonging to spin~$J$)
to the amputated scalar field $n$-point function (in this frame).
A tree has $J<2n_c{+}n_o$, and one loop means $J=2n_c{+}n_o$.
It turns out that the momentum dependence of all amplitudes occurs only
through the combinations
\be
k_i \wedge k_j \ :=\ \e_{\ad\bd}\;k_i^{+\ad}\,k_j^{+\bd} \quad.
\ee
For $n{=}3$, momentum conservation and $k_i^2{=}0$ entail 
$k_1{\wedge}k_2=k_2{\wedge}k_3=k_3{\wedge}k_1$.
Our present knowledge about $N{=}2$ string scattering amplitudes is
combined in the following table~\cite{DL2}:

\begin{center}
\begin{tabular}{|l|cccl|c|c|}
\hline		\vphantom{$\displaystyle\int$}
memo & $n_c$ & $n_o$ & $J$ & topology 
	& $g^J\,\tilde{A}^{(n_c,n_o)}_{J,J}$ & $\cl_{\rm int}(\Psi,\Phi)$ \\
\hline		\vphantom{$\displaystyle\int$}
$\langle ccc\rangle$ & 3 & 0 & 2 & sphere 
	& $\k\;(k_1 \wedge k_2)^2$         & $\Psi\,dd\Psi \wedge dd\Psi$ \\
		\vphantom{$\displaystyle\int$}
		     &   &   & 4 & disk, $\R P_2\!$ 
	& $\g\;\k\;(k_1 \wedge k_2)^4$     & $\Psi\,dddd\Psi \wedge dddd\Psi$\\
		\vphantom{$\displaystyle\int$}
		     &   &   & 6 & T,K,A,M 
	& $\e\;\k^3\;(k_1 \wedge k_2)^6$   & $\rightarrow$ one loop \\
		\vphantom{$\displaystyle\int$}
$\langle cco\rangle$ & 2 & 1 & 3 & disk
	& 0                                & --- \\
		\vphantom{$\displaystyle\int$}
		     &   &   & 5 & A, M
	& 0			           & $\rightarrow$ one loop \\
		\vphantom{$\displaystyle\int$}
$\langle coo\rangle$ & 1 & 2 & 2 & disk
	& $k^{AB}\,\k\;(k_1\wedge k_2)^2$  & $\Psi\,dd\Phi \wedge dd\Phi$ \\
		\vphantom{$\displaystyle\int$}
		     &   &   & 4 & A, M
	& $\zeta\;k^{AB}\,\k g^2\;(k_1 \wedge k_2)^4$ &$\rightarrow$ one loop\\
		\vphantom{$\displaystyle\int$}
$\langle ooo\rangle$ & 0 & 3 & 1 & disk
        & $f^{ABC}\,g\;(k_1 \wedge k_2)$   & $\Phi\,[d\Phi,\wedge\,d\Phi]$ \\
		\vphantom{$\displaystyle\int$}
                     &   &   & 3 & A, M
	& $\eta\;f^{ABC}\,g^3\;(k_1 \wedge k_2)^3\!$ &$\rightarrow$ one loop\\ 
\hline		\vphantom{$\displaystyle\int$}
$\langle c^{n\ge4}\rangle$ & $\!{\ge}4\!\!$ & 0 & $\!\!\!2n{-}4\!\!\!$ &sphere 
	& 0				   & --- \\
		\vphantom{$\displaystyle\int$}
$\langle o^{n\ge4}\rangle$ & 0 & $\!{\ge}4\!\!$ & $\!\!n{-}2\!\!$ & disk 
	& 0				   & --- \\
\hline
\end{tabular}
\end{center}

Here, T, K, A, and M stand for 
torus, Klein bottle, annulus, and M\"obius strip, respectively.
The coefficient~$\g$ is a finite constant of dimension cm$^4$, whereas the
one-loop coefficients $\e$, $\zeta$, and $\eta$ are infrared divergent.
Open string legs carry Chan-Paton labels~$A,B,\ldots$, which appear in
the Killing form~$k^{AB}$ and the structure constants~$f^{ABC}$ of the 
gauge group.
Beyond $n{=}3$, mixed open/closed tree amplitudes are expected to vanish
as well, a fact that has so far been shown only for 
$\langle ccco\rangle$ and $\langle cooo\rangle$~\cite{Mar}.

\bigskip\noindent
{\bf 4.\ \ Spacetime Dynamics.}
\smallskip

It is now straightforward to extract the (tree-level) vertices of the
spacetime field theory containing $\Psi$ and $\Phi$.
The contributions to $\cl_{\rm int}$ are listed in short-hand in the
table above, with the abbreviations
\be
d\ \ \cor\ \ \pa_{+\ad}\,,\;\pa_{+\bd}\,,\ldots \qquad{\rm and}\qquad
\wedge\ \ \cor\ \ \e^{\ad\bd}\,\e^{\gd\dd}\ldots \quad.
\ee
Since $\Phi$ is Lie-algebra valued, a trace is understood where it appears.
Apparently, four fundamental cubic vertices occur.
Those and the massless propagators must be used in Feynman diagrams
to evaluate higher $n$-point functions.
The mismatch of the field theory result with the corresponding string
amplitude is then cured by adding higher fundamental vertices to the
spacetime action.
The latter will make new contributions to even higher $n$-point functions,
so that an iterative process is started.
For the case at hand, however, it turns out that beyond $n{=}3$ 
all tree-level $n$-point functions based on the four cubic vertices
just vanish! Non-trivial kinematical identities in $2{+}2$ dimensions
ensure that the different scattering channels conspire and combine to zero.
Hence, no further vertices need to be added to the spacetime action in order
to reproduce the complete tree-level string scattering.

\medskip

If one chooses to have the couplings in front of the interaction terms,
a comparison of $\cl_{\rm int}$ with the kinetic terms 
$-\tr\sfrac12\Phi\square\Phi$ and $-\sfrac12\Psi\square\Psi$ reveals that
the scalars carry non-canonical length dimensions
$[\Phi]={\rm cm}^0$ and $[\Psi]={\rm cm}^1$
which necessitates dimensionful constants ($T$ and $T^2$) in front
of the action.
In total, the (tree-level) spacetime Lagrangian reads~\cite{DL2}
\bea\label{lag}
\cl_4\ &=&\ T\;\tr\,\Bigl\{ -\sfrac12\,\Phi\square\Phi 
+ \sfrac{g}6\,\Phi\,[d\Phi,\wedge\,d\Phi]
+ \sfrac{\k}2\,\Psi\,dd\Phi\wedge dd\Phi \Bigr\} \nonumber\\
&&+\ T^2\,\Bigl\{ -\sfrac12\,\Psi\square\Psi
+ \sfrac{\k}6\,\Psi\,dd\Psi\wedge dd\Psi
+ \g \sfrac{\k}6\,\Psi\,dddd\Psi\wedge dddd\Psi \Bigr\}
\eea
and displays a dimensional hierarchy:
To leading order ($T^2$) in the string tension, there is only the
basic closed-string vertex; to order $T$ the open-string sector joins;
and to order $T^2\g\sim T^0$ the unusual eight-derivative vertex comes in.
In the $T{\to}\infty$ limit of the $N{=}2$ string, the $O(T)$ and $O(1)$ 
terms are seen as ``stringy corrections'' to the $\Psi$ dynamics. 
Freezing the $\Psi$ dynamics, however, open strings ($\Phi$)
in the $\Psi$ background are dominant.
In view of the absence of massive string excitations, 
it is quite peculiar to observe a dependence on the string tension
for the effective field theory. Obviously, one must distinguish between
the full spacetime field theory and its ``field theory limit''.

\medskip

Since $\smallint\!d^4\!x\,\cl_4$ is a tree-level effective action,
its non-renormalizability need not concern us.  One can, however, 
improve on this drawback by using as alternative Lagrangian~\cite{CS}
\bea\label{cs}
\cl'_4\ &=&\ \!\tr\,\Bigl\{ -\tphi\square\Phi
+ \sfrac{g}2\,\tphi\,[d\Phi,\wedge\,d\Phi]
+ \k\,\tphi\,dd\Psi\wedge dd\Phi \Bigr\} \nonumber\\
&&+\ \Bigl\{ -\tpsi\square\Psi
+ \sfrac{\k}2\,\tpsi\,dd\Psi\wedge dd\Psi
+ \g \sfrac{\k}2\,\tpsi\,dddd\Psi\wedge dddd\Psi \Bigr\}
\eea
where Lagrange multiplier fields $\tphi$ and $\tpsi$ with
$[\tphi]={\rm cm}^{-2}$ and $[\tpsi]={\rm cm}^{-3}$
have been introduced to absorb the dimensionful constants.
Based on the phenomenon of ``ghost pictures'' in string theory,
there is a certain rationale for enlarging the number of fields
(even to infinity), although such fields do not represent new degrees
of freedom.
Alternatively in field theory, $\tphi$ and $\tpsi$ may be interpreted
as the other helicity state of the gluon and graviton, respectively.
The lagrangian \gl{cs} yields the same equations of motion (plus two more)
as the original one \gl{lag}. Yet, it is not only renormalizable
but even one-loop exact since powers of $\hbar$ go with inverse powers of
$\tphi$ or $\tpsi$.
Nevertheless, the off-shell structure of these actions is incomplete
unless a comparison with string amplitudes is made at the loop level.

\medskip

The tree-level string amplitudes only tell us the
field equations of motion,
\bea
\label{open}
&&-\square\Phi + \sfrac{g}2\,[d\Phi,\wedge\,d\Phi]
+ \k\,dd\Psi\wedge dd\Phi\ =\ 0 \\[1ex] 
\label{closed}
&&-\square\Psi + \sfrac{\k}2\,dd\Psi\wedge dd\Psi
+ \sfrac1T\,\sfrac{\k}2\,dd\Phi\wedge dd\Phi 
+ \g \sfrac{\k}2\,dddd\Psi\wedge dddd\Psi\ =\ 0
\eea
which I display more explicitly in the $T{\to}\infty$ limit:
\bea
\label{lez}
&&-\square\Phi^A  
+ \sfrac{g}2 f^{ABC}\,\p_+^{\ \ad}\Phi^B\,\p_{+\ad}\Phi^C
+ \k\,\p_+^{\ \ad}\p_+^{\ \bd}\Psi\,\p_{+\ad}\p_{+\bd}\Phi^A\ =\ 0 \\
\label{ple}
&&-\square\Psi 
+ \sfrac{\k}2\,\p_+^{\ \ad}\p_+^{\ \bd}\Psi\,\p_{+\ad}\p_{+\bd}\Psi\ =\ 0
\quad.
\eea
The $\Phi$-$\Psi$ coupling has disappeared from the $\Psi$
equation as a string correction, so that this set of equations is not
Lagrangian.
Hence, an action principle for the spacetime dynamics requires the string
extension of eqs. \gl{lez} and \gl{ple}. 

\bigskip\noindent
{\bf 5.\ \ Self-Duality Interpretation.}
\smallskip

The equations of motion \gl{lez} and \gl{ple} are known to mathematicians and 
integrable systems experts as the {\it curved Leznov equation\/}~\cite{Lez} 
and the {\it second Plebanski equation\/}~\cite{Ple}, respectively.  
They describe {\it self-dual Yang-Mills\/} coupled to {\it self-dual gravity\/}
in $2{+}2$ dimensional spacetime, in a particular gauge.
Let me demonstrate this connection, first in the flat limit $\k{\to}0$.

\medskip

Any two-form field strength $F=dA+A{\wedge}A$ decomposes into 
its self-dual and anti-self-dual part, $F=F_{(0,1)}+F_{(1,0)}$. 
The vanishing of $F_{(1,0)}$ is equivalent to the three component equations 
\be
F_{\a\b}\ =\ 
\p_{(\a}^{\ \gd}\,A_{\b)\gd}\ +\ g\,[A_\a^{\ \gd}\,,\,A_{\b\gd}]\ =\ 0 \quad.
\ee
These equations for the gauge connection~$A$ can be reduced to 
a single second-order equation for the Leznov prepotential $\Phi$ 
by going to the light-cone gauge 
\be
A_{+\gd}\ =\ 0 \quad.
\ee
This gauge incidentally selects a preferred spinor $v$ and thus breaks 
the `Lorentz invariance' as in \gl{break2}.
The remaining equations for $A_{-\gd}$ read
\be
0\ =\ \p_+^{\ \gd}\,A_{-\gd} \qquad{\rm and}\qquad
0\ =\ \p_-^{\ \gd}\,A_{-\gd} + \sfrac{g}2\,[A_-^{\ \gd}\,,\,A_{-\gd}]\quad.
\ee
The first of these is solved by introducing the prepotential $\Phi$ via
\be
A_{-\gd}\ =\ \p_{+\gd}\,\Phi
\ee
the use of which turns the second equation into
\be
0\ =\ -\square\Phi\;+\;\sfrac{g}2\,\p_+^{\ \gd}\Phi\,\p_{+\gd}\Phi 
\ee
which indeed is recognized as \gl{lez} at $\k{=}0$.
The field strength is recovered as
\be
F_{\gd\dd}\ =\ \p_{+\gd}\,\p_{+\dd}\,\Phi \quad.
\ee

\medskip

Let me turn to self-dual gravity. 
The curvature two-form $R=d\o+\o{\wedge}\o$ takes values in 
$sl(2,\R){+}sl(2,\R)'$, with generators denoted by $\Gamma^{\cdot\cdot}$. 
Setting to zero its $(1,0)$ half, in components
\be
0\ =\ R_{\a\b}\ =\ \e_{\a\g}\e_{\b\d}\,R\,\Gamma^{\g\d} 
+ C_{\a\b\g\d}\,\Gamma^{\g\d} + R_{\a\b\gd\dd}\,\Gamma^{\gd\dd} \quad,
\ee
eliminates all but the self-dual part of the Weyl tensor,
$R_{\ad\bd}=C_{\ad\bd\gd\dd}\,\Gamma^{\gd\dd}$.
In the light-cone gauge
\be
\o_{+\ad\gd\dd}\,\Gamma^{\gd\dd}\ =\ 0 \quad,
\ee
the ensueing equations for the spin connection~$\o$ simplify to
\be
\p_{+[\bd}\,\o_{-\ad]\gd\dd}=0 \ ,\quad
\o_{-[\ad\gd]\dd}=0 \ ,\quad
\o_{-\ad[\gd\dd]}=0 \ ,\quad
(D_{-[\ad}\,\o_{-\bd]})_{\gd\dd}=0 
\ee
where $D$ is the gravitationally covariant derivative (involving $\o$).
The first three of these equations are solved by introducing the 
prepotential~$\Psi$ via
\be
\o_{-\ad\gd\dd}\ =\ \p_{+\ad}\,\p_{+\gd}\,\p_{+\dd}\,\Psi
\ee
which allows one to rewrite the fourth equation as (a second derivative of)
\be
0\ =\ -\square\Psi\;+\;\sfrac{\k}2\,
\p_+^{\ \gd}\p_+^{\ \dd}\Psi\,\p_{+\gd}\p_{+\dd}\Psi
\ee
which really is identical to \gl{ple}.
The self-dual Weyl tensor emerges as
\be
C_{\ad\bd\gd\dd}\ =\ \p_{+\ad}\,\p_{+\bd}\,\p_{+\gd}\,\p_{+\dd}\,\Psi \quad.
\ee

\medskip

Since I implicitly used the vierbein when writing $F_{\a\b}{=}0$,
a curved metric background influences Yang-Mills self-duality.
As a result, all derivatives fully covariantize, in particular
$-\square\to D_-^{\ \ad}\p_{+\ad}$, which adds the term
\be
\k\,\p_+^{\ \ad}\p_+^{\ \bd}\Psi\,\p_{+\ad}\p_{+\bd}\Phi
\ee
to Leznov's equation when $\k{\neq}0$.
As expected, there is no back-reaction of the self-dual Yang-Mills 
energy-momentum tensor on the metric.

\medskip

Full agreement with the string effective field theory has been reached
for $T{\to}\infty$.
What about the ``stringy corrections'' in \gl{closed} to Plebanski's 
second equation~\gl{ple}?
Do they have a covariant interpretation in the self-dual framework?
Indeed, it is not hard to see that the last two terms in \gl{closed}
are proportional to
\be
\sfrac1T\,\k\,F \wedge F \qquad{\rm and}\qquad 
\g\,\k\,R \wedge R \quad{\rm or}\quad \g\,\k\,C \wedge C \quad,
\ee
respectively. Yet, since two derivatives are needed to move from 
Plebanski's to the self-duality equation, the modification of the latter
should schematically look like
\be
R_{(1,0)}\ \sim\ \k\,dd\,(a\,F \wedge F + b\,R \wedge R)
\ee
but I don't know how to write out these derivatives.
Let me close this part by remarking that other formulations of self-dual
Yang-Mills plus gravity can be reached from $N{=}2$ strings. Had I chosen
to average over the cotangent structures ($\th$) in \gl{amp} rather than
taken $\th{=}0$, scalar field dynamics governed by Yang's~\cite{Yang} and 
Plebanski's first equation~\cite{Ple} would have emerged. 
These equations again describe self-dual Yang-Mills and self-dual gravity, 
albeit in a different gauge.

\bigskip\noindent
{\bf 6.\ \ Infinite Symmetry from Pictures.}
\smallskip

Self-dual field theories are integrable.
It is therefore natural to search for a corresponding integrable structure
in $N{=}2$ strings. Let me concentrate briefly on the closed-string sector,
i.e. self-dual gravity as parametrized by Plebanski's second equation.

\medskip

It is known that self-dual gravity possesses an infinity class of non-local
symmetries, which form a subgroup of the loop group 
\be\label{sym}
\LSDiff^{\C}(M)\ =\  C^\infty (S^1,\SDiff^{\C}(M))
\ee
of area-preserving diffeomorphisms of the spacetime manifold~$M$.
They may be used to generate new solutions of Plebanski's equation
from old ones. 
General symmetries $\d\!\Psi$ of eq.~\gl{ple} satisfy
\be
-\square\,\d\!\Psi + \k\,dd\Psi \wedge dd\,\d\!\Psi\ =\ 0
\ee
and correspond to shifts in the moduli parametrizing the solution space.

\medskip

An infinite hierarchy
of {\it commuting\/} symmetry flows is generated by the maximal abelian
subalgebra $\C^2\otimes\C[\l]$ of \gl{sym}, with generators $\{\p^n_a\}$.
It is used to parametrize the solution space to Plebanski's equation
by coordinates $t={(t^a_n)}={(t^a_{-1}, t^a_0, t^a_1, ... )}$,
where $(t^a_{-1}, t^a_0)=(x^a, \e^a_\ba \bx^\ba)$.
Starting from the obvious symmetry under spacetime translation,
one can recursively solve the equations of the hierarchy and recover
the dependence on the moduli.
These non-local abelian symmetries are best described in momentum space,
where their linearization takes the form~\cite{JLP}
\be\label{lin}
\d_n^a\Psi(k)\ =\ k^a\,h(k)^{-n}\,\Psi(k) 
\quad,\qquad n=0,1,2,\ldots \quad,\qquad a=1,2 \quad,
\ee
with $k{\cdot}\bar{k}=0$ and the momentum phase
\be
h(k)\ =\ {k^1}/{\bar{k}^{\bar{2}}}\ =\ {k^2}/{\bar{k}^{\bar{1}}}\ =\
(h(k)^{-1})^* \quad.
\ee

\medskip

It is possible to recover such symmetries from first-quantized string theory.
However, in our single-string Fock space symmetries can only act linearly,
\be
\d\,|{\rm phys}\rangle\ =\ |\Lambda\rangle + A\,|{\rm phys}\rangle\ \quad.
\ee
To search for global symmetries which are unbroken 
($|\Lambda\rangle{=}0$) in the given background,
the proper tool is the cohomology of the BRST operator.
More precisely, all conserved charges reside in the BRST cohomology
at $k=0$ and at ghost number $gh=gh_{\rm phys}{-}1$.
This task has been solved~\cite{JLP,JL1} for the closed $N{=}2$ string, 
with the finding of an infinity of commuting symmetry charges $A$, 
constructed from the so-called ground ring. These charges are labeled by
\be
a\in\{1,2\}\ ,\quad
j,\tj\in\{ 0,\sfrac12,1,\sfrac32,\ldots\}\ ,\quad
m\in\{-j,\ldots,j\}\ ,\quad
\tm\in\{-\tj,\ldots,\tj\}
\ee
and act on physical states as follows,
\be
A^a_{j,m;\tj,\tm}\,|{\rm phys},k\rangle\ =\
k^a\,h(k)^{-(j-m)-(\tj-\tm)}\,|{\rm phys}',k\rangle
\ee
where $|{\rm phys}',k\rangle$ represents the same physical state as 
$|{\rm phys},k\rangle$.
Comparison with \gl{lin} yields an obvious match!
Moreover, it is quite satisfying to realize~\cite{JLP} that the 
Ward identites based on these symmetries are strong enough to enforce 
\be
k_i\cdot\bar{k}_j+\bar{k}_i\cdot k_j\ =\ 0 \qquad \forall\ i,j
\ee
for non-zero tree amplitudes.
This condition kills all trees beyond the three-point function,
in accordance with the explicit computations.

\medskip

Here, I would like to digress briefly to explain the picture 
phenomenon~\cite{FMS} as the origin of the newfound string symmetries.
In the BRST framework for local worldsheet supersymmetry, one introduces
first-order {\it commuting\/} ghosts $\g$ and anti-ghosts $\b$,
whose Fourier modes obey
\be
[ \g_r\,,\,\b_s ]\ =\ \d_{r+s,0} \qquad{\rm for}\quad
r,s\in\Z{+}\sfrac12 \quad {\rm in\ the\ Ramond\ sector}\quad.
\ee 
The separation of the ghost and anti-ghost modes into creation and
annihilation operators defines a ghost vacuum. There is a freedom
of choice, parametrized by an integer~$\pi$, in
\be
\g_r |\pi\rangle\ =\ 0 \quad{\rm for}\quad r\ge\sfrac12{-}\pi
\qquad{\rm and}\qquad
\b_r |\pi\rangle\ =\ 0 \quad{\rm for}\quad r\ge\sfrac12{+}\pi
\quad.
\ee
Due to the commuting nature of the $(\g,\b)$ ghosts,
different vacua $|\pi\rangle$ (and the Fock spaces built over them) 
are formally connected only through the action
of {\it distributions\/} like $\d(\g_r)$ in ghost/antighost modes
or by extending the Fock space via bosonization~\cite{FMS,Ver}.
For example, one may check that
\be
|{-}1\rangle\ =\ \d(\g_{\frac12})|0\rangle 
\qquad{\rm and}\qquad
|0\rangle\ =\ \d(\b_{-\frac12})|{-}1\rangle \quad.
\ee
Hence, different pictures are isomorphic but not identical.
At this stage, one may wonder if this picture degeneracy is relevant at all.
Why not just choose one of these pictures for good, 
say the one with the canonical value~$\pi{=}-1$?
It was recently discovered~\cite{JL1}, however, that the BRST cohomology 
for $k{=}0$ depends on the picture number!~\footnote{
Actually, this even happens in the RR sector of the $N{=}1$ string~\cite{BZ}.}
Therefore, ghost pictures, being conjugate to the ``higher time'' coordinates
in the moduli space of self-dual metrics, have a physical significance.

\bigskip\noindent
{\bf 7.\ \ Beyond Tree Level.}
\smallskip

In the absence of a Lorentz covariant action, 
there is no obvious recipe how to covariantly quantize 
$2{+}2$ dimensional self-dual field theories. 
Of course, one may naively take Plebanski's, Leznov's, Yang's or other 
actions off-shell, but the result is expected to depend on the chosen gauge.
Yet, at one loop there seems to be agreement~\cite{CS}.
More explicitly for self-dual Yang-Mills, 
the one-loop $n$-point function for the negative-helicity
gauge field $A_{\_}:=\c^\ad A_{-\ad}=\c^\ad\p_{+\ad}\Phi$ 
with polarization spinor $\c^\ad$ has been evaluated to be~\cite{MHV}
\be
\langle A_{\_}(k_1)\ldots A_{\_}(k_n)\rangle\ \sim\
\sum_{{ijkl\ {\rm cyclic}\atop\in\{1,2,\ldots,n\}}}
\frac{[ij]\langle jk\rangle[kl]\langle li\rangle}
     {\langle12\rangle\langle23\rangle\ldots\langle n1\rangle}
\qquad{\rm for}\quad n\ge4
\ee
where each lightlike momentum has been split in two momentum spinors,
$k^{\a\ad}=\l^\a\c^\ad$, which are then combined in
\be\label{loop}
[ij]\ :=\ \e_{\ad\bd}\,\c_i^\ad\,\c_j^\bd
\qquad{\rm and}\qquad
\langle ij\rangle\ :=\ \e_{\a\b}\,\l_i^\a\,\l_j^\b \quad.
\ee
The fact that this amplitude has only two-particle poles and no cuts 
in the complex momentum plane
is remarkable but consistent with the vanishing of the tree-level graphs.
It is also infrared finite (the $n{=}3$ result is not).
Furthermore, after Wick rotating from $2{+}2$ to $3{+}1$ dimensions,
\gl{loop} is also the leading-color QCD result for this so-called
``maximally helicity violating'' configuration (i.e. $A_-$ legs only)!
This has led to the conjecture that all such (MHV) amplitudes in QCD are
(to leading color) given by the self-dual sector alone.
At tree-level, such a statement is empty since the only non-vanishing
amplitudes ($n{=}3$) cannot be continued to Minkowski space.
The obvious kinematical difference is that the presence of null planes 
in $\R^{2,2}$ allows a massless particle to decay into two, 
which in $\R^{3,1}$ is possible only in the collinear limit.

\medskip

$N{=}2$ strings should provide a natural off-shell extension of (gauge-fixed)
self-dual field theories, since critical string theories are 
per construction quantum consistent. 
As I have demonstrated, $N{=}2$ strings classically agree (at $T{\to}\infty$)
with those gauge-fixed field theories.
However, this does not mean that they produce the same loop amplitudes.
Like for higher tree-level $n$-point functions, one must compare the one-loop
diagrams from the cubic action \gl{lag} or \gl{cs} with the corresponding
one-loop string amplitude and interpret the difference, if any,
as a further correction to the (classical) effective spacetime action.
This is the reason for the ``one loop'' entries in my table of amplitudes. 
Unfortunately, only three-string scattering has been evaluated at one loop
and found to be infrared divergent~\cite{BGI}, just like the one-loop 
three-point functions for Leznov's or Plebanski's second prepotential.
Clearly, the crucial test is the computation of an $N{=}2$ string
one-loop four-point amplitude, which may (after Wick rotation) be compared
directly with \gl{loop}.
This rather technical task is under way, with indications that the result
may be infrared finite.
If agreement between self-dual field theories and $N{=}2$ strings is found
also at the quantum level, the latter can be employed to calculate
entirely left-handed multi-gluon scattering in QCD 
(and analogous multi-graviton scattering in four dimensions).

\medskip

The frontier should be pushed back in two (related) directions.
First, we need more explicit string loop results, perhaps with the
help of Ward identities associated to the infinite set of symmetries.
Second, more general spacetime (plus gauge field) backgrounds ought to
be investigated. More concretely, compactifications, the spectrum and
charges of D-branes, possible dualities, and constant antisymmetric tensor
backgrounds are waiting to be worked out.


\noindent


\end{document}